\documentclass[aps,prb,twocolumn]{revtex4}
\usepackage{amsfonts}
\usepackage{amsmath}
\usepackage{graphicx}
\usepackage{dsfont}
\usepackage{diagbox}
\usepackage{array}
\usepackage{amssymb}
\usepackage{bbm}
\usepackage{float}
\usepackage{url}
\usepackage{hyperref}
\usepackage{booktabs}
\usepackage{xcolor}

\begin{document}

\title{Tensor network approach to the fully frustrated XY model on a kagome lattice 
with a fractional vortex-antivortex pairing transition}
\author{Feng-Feng Song$^{1}$ and Guang-Ming Zhang$^{1,2,3}$}
\email{gmzhang@tsinghua.edu.cn}
\affiliation{$^{1}$State Key Laboratory of Low-Dimensional Quantum Physics and Department
of Physics, Tsinghua University, Beijing 100084, China\\
$^{2}$Collaborative Innovation Center of Quantum Matter, Beijing 100084,
China\\
$^{3}$Frontier Science Center for Quantum Information, Beijing 100084, China}
\date{\today}

\begin{abstract}
We have developed a tensor network approach to the two-dimensional fully 
frustrated classical XY spin model on the kagome lattice, and clarified the 
nature of the possible phase transitions of various topological excitations. 
We find that the standard tensor network representation for the partition 
function does not work due to the strong frustrations in the low temperature 
limit. To avoid the direct truncation of the Boltzmann weight, based on the 
duality transformation, we introduce a new representation to build the tensor 
network with local tensors lying on the centers of the elementary triangles 
of the kagome lattice. Then the partition function is expressed as a product 
of one-dimensional transfer matrix operators, whose eigen-equation can be 
solved by the variational uniform matrix product state algorithm accurately. 
The singularity of the entanglement entropy for the one-dimensional quantum 
operator provides a stringent criterion for the possible phase transitions. 
Through a systematic numerical analysis of thermodynamic properties and 
correlation functions in the thermodynamic limit, we prove that the model 
exhibits a single Berezinskii-Kosterlitz-Thouless phase transition only, 
which is driven by the unbinding of $1/3$ fractional vortex-antivortex pairs 
determined at $T_{c}\simeq 0.075J_{1}$ accurately. The absence of long-range 
order of chirality or quasi-long range order of integer vortices has been 
verified in the whole finite temperature range. Thus the long-standing 
controversy about the phase transitions in this fully frustrated XY model on 
the kagome lattice is solved rigorously, which provides a plausible way to 
understand the charge-6e superconducting phase observed experimentally in the 
two-dimensional kagome superconductors.
\end{abstract}

\maketitle

\section{Introduction.}

Various two dimensional (2D) kagome lattice models have attracted a lot of
interest in the study of the interplay between band topology, geometric
frustrations and strong correlations in the past decades. Due to the special
lattice geometry with corner-sharing triangles, the kagome lattice features
exotic electronic structures, such as flat bands, van Hove singularity and
non-trivial topology of Dirac cones\cite%
{Mielke_1991,Neupert_2022,Yin_2022,Jiang_2023}. The strongly geometric
frustrations can induce massive degenerate ground states in kagome spin
lattice models, providing one of the most promising platforms to realize
quantum spin liquids\cite{Balents_2010,Norman_2016,Zhou_2017,Broholm_2020}.
In addition to these intriguing phenomena, the unusual kagome
superconductivity (SC) has recently been intensively investigated in the
family of quasi-2D materials AV${}_{3}$Sb${}_{5}$ (A=K, Rb, Cs) (Ref. \cite%
{Ortiz_2019,Ortiz_2020,Ortiz_2021,Jiang_2021,Liang_2021}). One of the
exciting discoveries is the possible vestigial charge-6e SC around the
superconducting transition\cite{Ge_2022}.

Different from the conventional SC described by the
Bardeen-Cooper-Schrieffer theory as condensation of charge-2e Cooper pairs,
the charge-4e or -6e SC can emerge as a vestigial higher-order condensation
of bound states of electron sextets above the critical temperature of
charge-2e SC (Ref.\cite{Kivelson_1990,Ropke_1998,Wu_2005,Agterberg_2008,Berg_2009,Ko_2009,
Herland_2010,Jiang_2017,Fernandes_2021,Jian_2021,Song_2022_2,Agterberg_2011,Struck_2011,You_2012}). 
It seems that the charge-6e SC has a close relation to certain types of pair 
density wave states on a hexagonal lattice\cite{Chen_2021}. The most important 
feature of the charge-6e SC is characterized by the fractional magnetic flux 
quantization of $h/6e$. Despite the latest experimental progress, many basic 
properties of charge-6e SC remain less well understood. Some theoretical 
studies\cite{Agterberg_2011,Zhou_2022,Pan_2022} have suggested that the 
charge-6e SC may be resulted from the fluctuations of pair density wave order. 
Nevertheless, the experimental realization of such charge-6e SC has not been 
verified so far.

Despite the lack of direct physical origin and implication of these
remarkable observations, the crucial ingredient of the underlying geometry
frustrations provides an insightful perspective to understand the charge-6e
SC. The exotic quantum phenomena of charge-6e SC can naturally be caused by
the effect of strong frustrations of the kagome geometry. At the
phenomenology level, a complex SC order parameter can be expressed as $|\psi
(\vec{r})|e^{i\theta (\vec{r})}$. When the amplitude fluctuations of $|\psi (%
\vec{r})|$ is frozen, the SC transition is governed by the phase
fluctuations of the Cooper pairs, and the essential physics is described by
a classical \textit{ferromagnetic} XY spin model\cite%
{Beasley_1979,Emery-Kivelson,Carlson-1999,Li-2007}. However, in the presence
of an external magnetic field, the ferromagnetic coupling between the
nearest-neighbor (NN) XY spins may be tuned to be antiferromagnetic when
each triangle of the kagome lattice has a $\pi $ flux\cite{Rzchowski_1997,Park_2001}, 
which is the fully frustrated XY spin model. Such a model exhibits a large 
ground state degeneracy and fractional topological excitations because of the
interference on the underlying lattice topology\cite%
{Harris_1992,Rzchowski_1997,Cherepanov_2001,Park_2001,Korshunov_2002,Andreanov_2020}.

For the 2D antiferromagnetic XY kagome lattice model, apart from the global
$U(1)$ spin rotational degrees of freedom, the frustration in each elementary 
triangle can induce the chiral degrees of freedom. To minimize the energy of each
triangular plaquette, the orientation of the corresponding three XY spins
should be different from each other by an angle of $2\pi /3$. In this way,
when these three spins rotate clockwise or anti-clockwise, each plaquette
can be ascribed a positive or negative chirality, displayed in Fig~\ref%
{fig:kgm_H}. In addition to the degeneracy associated with $U(1)$ symmetry, 
the ground state has a finite residual entropy of $0.126k_{B}$ per site, 
similar to the $3$-state antiferromagnetic Potts model\cite{Baxter_1970,Huse_1992}. 
At low temperatures, the fluctuations of chirality are so strong that the phase
angle between two spins separated by a long distance can freely change by $%
\pm 2\pi /3$. This uncertainty in the phase difference increases with
distance and destroys the quasi-long-range order (quasi-LRO) in the XY
spins. The absence of the quasi-LRO of the XY spins $\exp (i\theta )$ implies 
that the phase coherence of the charge-2e SC is absent. 

\begin{figure}[tbp]
\centering
\includegraphics[width=\linewidth]{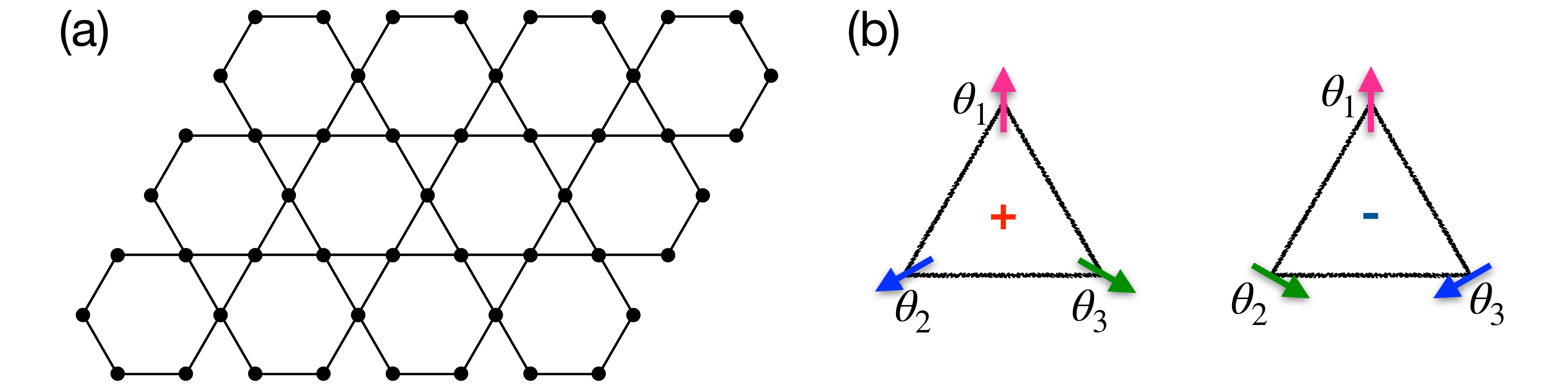}
\caption{ (a) The kagome lattice with anti-ferromagnetic nearest neighbor
couplings. (b) Triangular plaquette with $+$ and $-$ chirlaity.}
\label{fig:kgm_H}
\end{figure}

Although many theoretical conjectures were proposed, there has not yet had a
general consensus on the existence of the phase with LRO arrangement of
chirality at low temperatures. Some studies\cite{Harris_1992,Korshunov_2002}
argued that the spin wave fluctuations favor the LRO of the
anti-ferromagnetic arrangement of chirality for the neighboring triangular
plaquettes and the phase coherence between integer vortices survives, while
some other works\cite%
{Rzchowski_1997,Cherepanov_2001,Park_2001,Andreanov_2020} suggested that the
chirality is always disordered corresponding to the suppression of the
conventional charge-2e SC and the lowest-order quantity displaying quasi-LRO
is given by the $\exp(i3\theta)$ variables. Due to the presence of strong
frustrations and massive degeneracy of the ground state, it reminds a
challenging problem to clarify the low temperature phases of the
anti-ferromagnetic XY model on the kagome lattice. The reason for the
numerical difficulty is that the excitations of different kinds of
topological excitations (integer vortices, $1/3$ fractional vortices, domain
walls, and kinks on the domain walls) occur close to each other at rather
low transition temperatures of nearly $1/11$ of the usual non-frustrated 
Berezinskii-Kosterlitz-Thouless (BKT) transition\cite{Rzchowski_1997}. And 
the traditional sampling methods often suffer from the finite size effects 
and a critical slowing down when approaching to the low temperature 
phase\cite{Swendsen_1987,Wolff_1989,Andreanov_2020}.

Fortunately, the recent developments in tensor network (TN) methods have
shed new light on the numerical study of 2D classical frustrated spin systems%
\cite{Vanderstraeten_2018,Vanhecke_2021,Song_2022,Colbois_2022}. It is found
that the construction of the TN representations of the partition functions
should be carried out with special care to encode the underlying physics of
all the ground-state configurations at the level of local tensors. The
convergence problem of the contraction algorithms of TN can be overcome by
suitable Hamiltonian tessellations or split of $U(1)$ spins. In this work,
we develop a peculiar TN approach to numerically solve the anti-ferromagnetic 
XY model on kagome lattices. Different from the discrete cases of $Z_{2}$ 
spins\cite{Vanhecke_2021,Colbois_2022} or the fully frustrated XY spin model 
on a square lattice with a chiral LRO ground state\cite{Song_2022}, the
choice of the TN representations for the fully frustrated kagome lattice
model does not affect the convergence of the contraction algorithms, but it
does lead to incorrect contraction values at low temperatures. We find that
the standard formulation is not always a good option, because the finite
truncation of the Boltzmann weights fails to represent the massive
degeneracy of the ground states. Here we have to introduce a special
construction strategy based on the duality transformations\cite%
{Jose_1977,Knops_1977,Cherepanov_2001}, and then the partition function is
transformed into an infinite 2D TN with local tensors lying on the centers
of the elementary triangles of the original kagome lattice, which can be
efficiently contracted by a recently proposed TN algorithm under optimal
variational principles\cite%
{Zauner_Stauber_2018,Fishman_2018,Vanderstraeten_2019}.

Once the proper implementation of TN representation is achieved, it can be
expressed in terms of a product of 1D transfer matrix operator. The
singularity of the entanglement entropy of this 1D quantum transfer operator
can be used to determine various phase transitions with good accuracy\cite%
{Haegeman-Verstraete2017}. The distinct advantage of the TN method is that a
stringent criterion can be used to distinguish various phase transitions
without identifying any order parameter \textit{a prior}. From the
perspective of the quantum entanglement, we determine the phase structure of
the anti-ferromagnetic XY model on the kagome lattices with clear evidence
that only a single phase transition takes place at $T_{c}\simeq 0.075J_{1}$.
From the analysis of the thermodynamic properties and the behavior of XY
spin correlation functions, we demonstrate the phase transition belongs to
the BKT universality class\cite%
{Berezinsky_1970,Kosterlitz_1973,Kosterlitz_1974} driven by the unbinding of
$1/3$ fractional vortex-antivortex pairs. In the absence of LRO in chirality, 
the low temperature phase is characterized by a quasi-LRO in the spin variable 
$\exp(i3\theta )$, while the phase coherence of integer vortices is destroyed 
due to the exponential decaying of the $\exp (i\theta )$ correlation functions.
The clarification of such a phase gives rise to a prototype example, where
the conventional charge-2e SC of the Cooper pairs is suppressed, but an
unusual form of charge-6e SC of \textquotedblleft Cooper
sextuples\textquotedblright\ survives.

The rest of the paper is organized as follows. In Sec. II, we give an
introduction of the 2D fully frustrated XY model on the kagome lattice and
the possible phase transitions. In Sec. III, we develop a general framework
of the TN method for this fully frustrated model based on the crucial
duality transformation. In Sec. IV, we present the numerical results for the
determination of the finite temperature phase diagram. Finally in Sec. V, we
give our conclusions and future extensions of our work.

\section{Model Hamiltonian and topological excitations}

We start with the effective Ginzburg-Landau (GL) free-energy density of
superconductivity in an external gauge field
\begin{equation}
\mathcal{F}_{GL}=a|\psi |^{2}+\frac{b}{2}|\psi |^{4}+\frac{1}{2m^{\ast }}%
\left\vert \left( \frac{\hbar }{i}\nabla -\frac{e^{\ast }}{c}\vec{A}\right)
\psi \right\vert ^{2},
\end{equation}%
where $a$, $b$ are phenomenological expansion coefficients dependent on the
materials, $m^{\ast }=2m_{e}$ and $e^{\ast }=2e$ are the effective mass and
the charge of Cooper pairs, and the energy of the inductive fields $\vec{B}%
=\nabla \times \vec{A}$ produced by the supercurrents are ignored\cite%
{Park_2001}. The complex order parameter $\psi (\vec{r})=\langle c_{\uparrow
}^{\dagger }(\vec{r})c_{\downarrow }^{\dagger }(\vec{r})\rangle $ can be
regarded as a wave function for the Cooper pairs and expressed as $\psi (%
\vec{r})=|\psi (\vec{r})|\mathrm{e}^{i\theta (\vec{r})}$. Since the
fluctuations of $|\psi (\vec{r})|$ can be ignored in the low temperature
regime, we reduce the above GL free energy to a classical XY spin model on a
2D kagome lattice with the model Hamiltonian\cite{Teitel_1983}
\begin{equation}
H=-J_{1}\sum_{\langle ij\rangle }\cos \left( \theta _{i}-\theta
_{j}+A_{ij}\right) ,  \label{eq:ffxy}
\end{equation}%
where $J_{1}=\frac{n_{s}\hbar ^{2}}{m^{\ast }}$ is the coupling strength
with the superfluid density fixed as $n_{s}=|\psi (\vec{r})|^{2}$, $i$ and $%
j $ enumerate the lattice sites, $\theta _{i}$ can be regarded as classical
XY spins associated with each lattice site and the summation is over all
pairs of the nearest neighbors. The frustration is induced by the gauge
field defined on the lattice bond satisfying $A_{ij}=-A_{ji}$. The gauge
field is related to the vector potential by $A_{ij}=\frac{2\pi }{\Phi _{0}}%
\int_{\vec{r}_{i}}^{\vec{r}_{j}}d\vec{l}\cdot \vec{A}$, where $\Phi _{0}=%
\frac{h}{2e}$ is the flux quantum.

In the absence of external magnetic field, the effective coupling is
ferromagnetic and results in a conventional BKT phase transition to the low
temperature state with power-law correlations of the XY spins, corresponding
to the charge-2e SC. The corresponding physics of this transition has been
well-established, which does not depend on the specifics of the lattice
structure. To introduce strong frustrations into this model, we should have
a $Z_{2}$ gauge field (one-half quantum flux per plaquette) $\sum_{\langle
i,j\rangle \in \triangle }A_{ij}\equiv \pi \pmod{2\pi}$, where the summation
is taken around the perimeter of each elementary triangle. Note that the direct
physical origin of the gauge field in realistic kagome SC materials is still
unclear, which may be resulted from the external magnetic fields\cite%
{Teitel_1983,Rzchowski_1997,Park_2001}.

\begin{figure}[tbp]
\centering
\includegraphics[width=\linewidth]{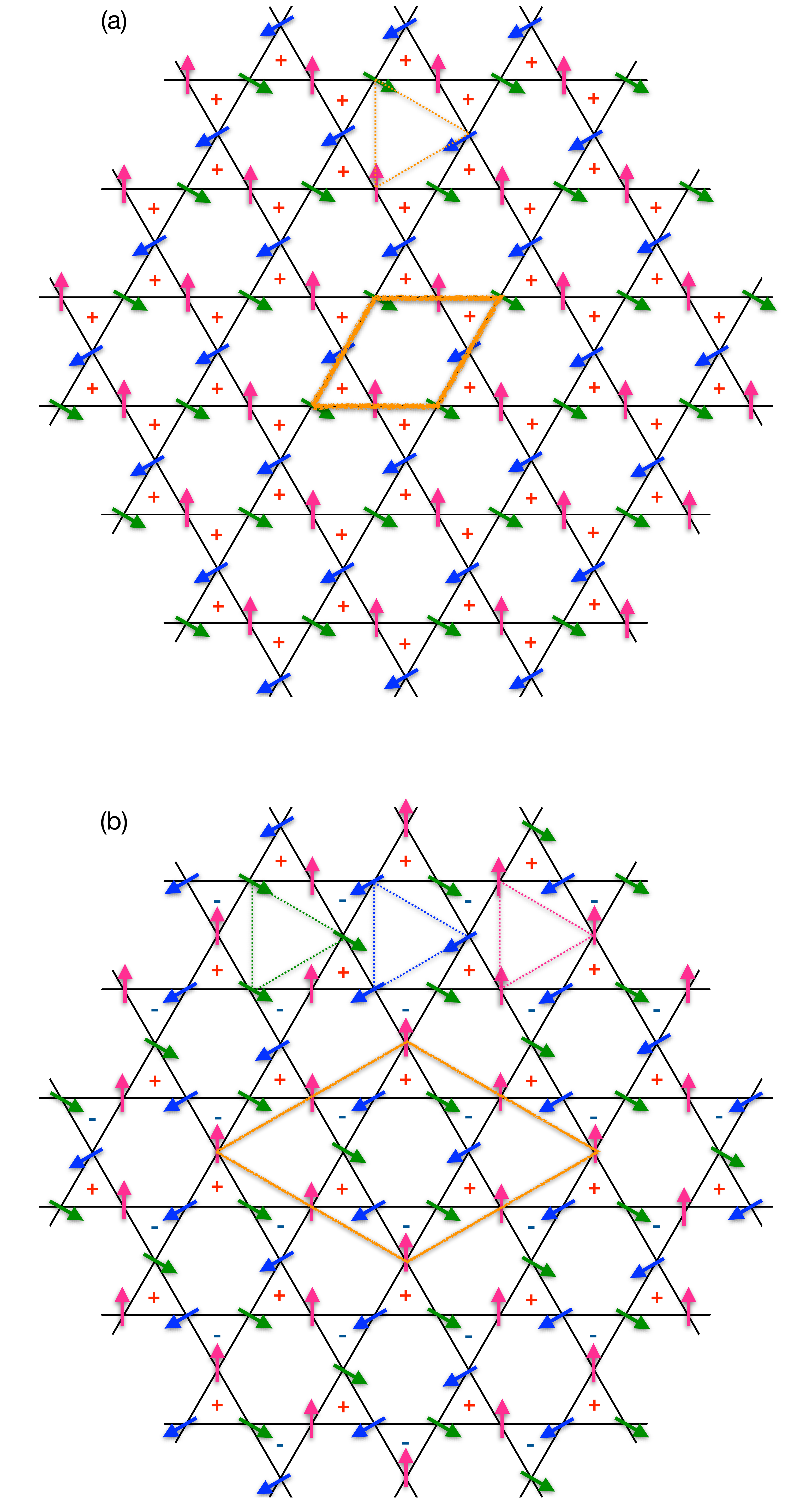}
\caption{ Two periodic patterns of chirality. Each pattern has another $Z_{2}$
degenerate state by switching the positive and negative chirality. The translational
invariant units are represented by orange parallelograms, respectively.
(a) One of the two $q=0$ patterns with a ferromagnetic arrangement of
chiralities. (b) One of the two $\protect\sqrt{3}\times \protect\sqrt{3}$
patterns with a regular alternation of positive and negative chiralities.}
\label{fig:ground}
\end{figure}

Then the model Hamiltonian of the anti-ferromagnetic XY model on the kagome
lattices is given by
\begin{equation}
H=J_{1}\sum_{\langle i,j\rangle }\cos (\theta _{i}-\theta _{j}),
\end{equation}%
where each elementary triangular plaquette is frustrated due to the
anti-ferromagnetic interactions. To achieve the minimum of the energy, the
angle between each pair of nearest neighbor spins should be $\pm 2\pi /3$.
In addition to the global $U(1)$ rotation of all spins related to the global
invariance of the Hamiltonian like the 2D classical XY model, the elementary
triangular plaquette can be characterized by the chirality $\sigma =\pm 1$.
The positive and negative chiralities of the plaquette are displayed in Fig~%
\ref{fig:kgm_H} (b), corresponding to the anti-clockwise and clockwise
rotation of the spins, respectively. The ground states of the
anti-ferromagnetic XY model on a kagome lattice present a massive accidental
degeneracy described by the $3$-state anti-ferromagnetic Potts model, which
can also be mapped onto a solid-on-solid model at its roughening transition%
\cite{Huse_1992}.

The accidental degeneracy of the ground state can be reduced to $U(1)\times
Z_{2}$ in the presence of next-to-nearest neighbors (NNN) interactions:
\begin{equation}
H=J_{1}\sum_{\langle i,j\rangle }\cos (\theta _{i}-\theta
_{j})+J_{2}\sum_{\langle \langle i,j\rangle \rangle }\cos (\theta
_{i}-\theta _{j}),  \label{eq:NNN_H}
\end{equation}%
where $J_{2}$ is the coupling between NNN sites denoted by $\langle \langle
i,j\rangle \rangle $. For the anti-ferromagnetic NNN interaction ($J_{2}>0$%
), the spin pattern of the ground state has the same translational period as
the usual kagome lattices, which is called the \textquotedblleft $q=0$
state\textquotedblright\ shown in Fig~\ref{fig:ground} (a). For the
ferromagnetic NNN interactions ($J_{2}<0$), the ground state of the spin
pattern is called the \textquotedblleft $\sqrt{3}\times \sqrt{3}$
state\textquotedblright\ with a regular alternation of positive and negative
chiralities. Such a pattern has a translational invariant unit including
three hexagons with a linear dimension $\sqrt{3}$ times larger than the unit
cell of the original kagome lattice as displayed in Fig~\ref{fig:ground} (b).

\begin{figure}[tbp]
\centering
\includegraphics[width=\linewidth]{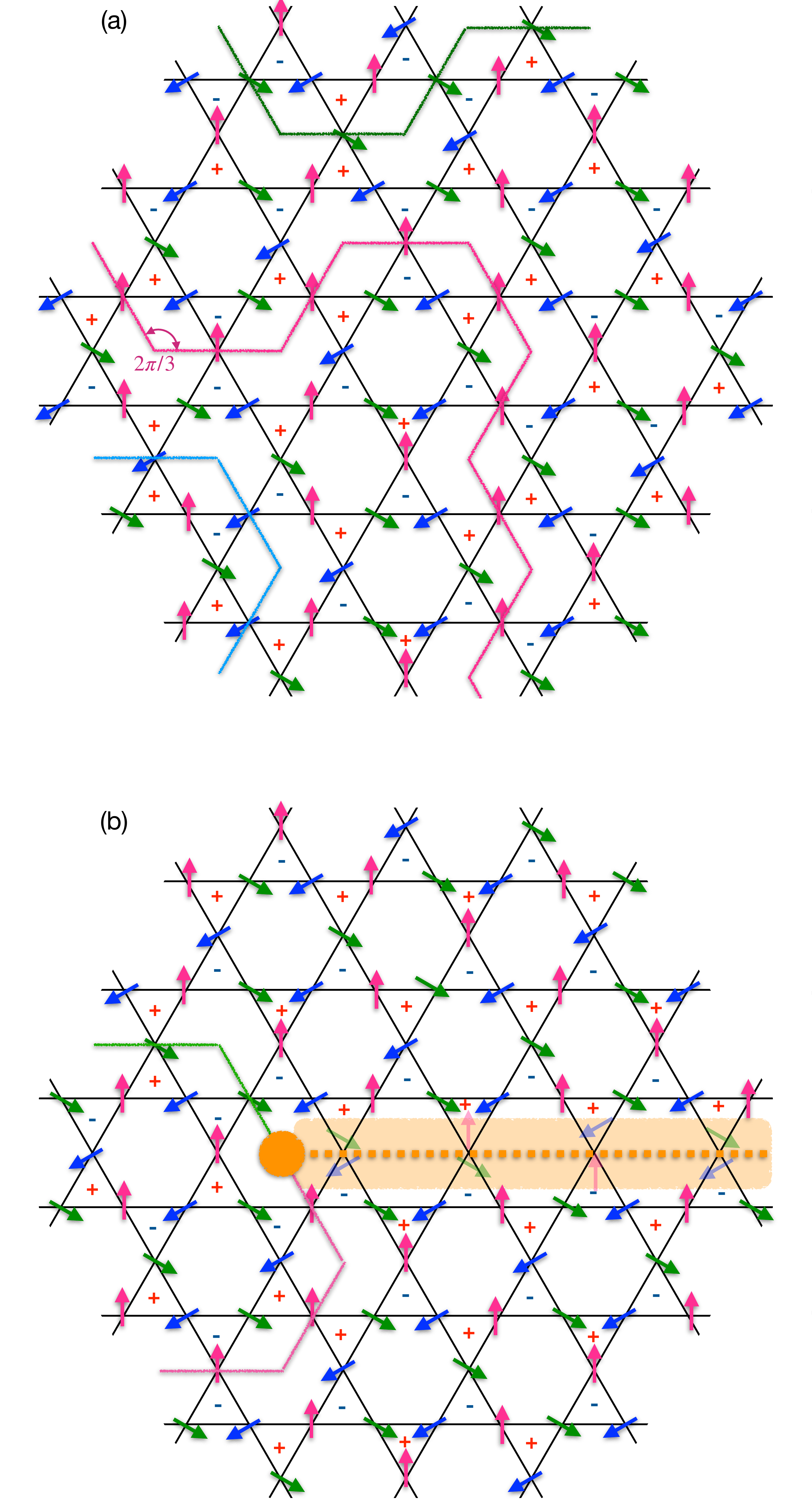}
\caption{ (a) Zero-energy domain walls (green, red and blue lines)
separating two triangular plaquettes with the same chirality belonging to
different $\protect\sqrt{3}\times \protect\sqrt{3}$ states. (b) Fractional
vortices (orange circle) appear at every point where elementary links
forming a domain wall meet each other at a wrong angle not equal to $2%
\protect\pi /3$. The kink introduces a string (dotted line) with a
discrepancy of $2\protect\pi /3$ starting from the fractional vortex and
terminating at another fractional vortex (not shown).}
\label{fig:excitations}
\end{figure}

Although extensive studies were carried out for this anti-ferromagnetic XY
model on 2D kagome lattices in the past decades, the nature of the phase
transitions in the frustrated model is still controversial \cite%
{Harris_1992,Rzchowski_1997,Cherepanov_2001,Park_2001,Korshunov_2002,Andreanov_2020}%
. The focus of the problem is whether the degeneracy of the chiralities
could be lifted strongly enough to drive the system into a phase with LRO
arrangement of chirality. Some studies\cite{Harris_1992,Korshunov_2002}
proposed that the spin wave fluctuations would favor the LRO of the
anti-ferromagnetic arrangement of chirality for the neighboring triangular
plaquettes ($\sqrt{3}\times \sqrt{3}$ state) at low temperatures through a
mechanism called \textquotedblleft order by disorder\textquotedblright . Due
to the $U(1)\times Z_{2}$ degeneracy, two kinds of topological excitations
are expected to exist: (i) linear defects as the domain walls separating two
ground states of different chirality patterns, (ii) point-like defects as
vortices or anti-vortices which destroy the $U(1)$ phase coherence. Fig.~\ref%
{fig:excitations} (a) displays the zero-energy domain walls separating the
different $\sqrt{3}\times \sqrt{3}$ patterns. Each segment of the domain
wall separates two triangular plaquettes with the same chirality and forms
an angle of $2\pi /3$ between each other. Moreover, possible fractional
vortices with topological charges of $\pm 1/3$ can exist on the kinks on the
domain walls, where elementary links meet each other at a wrong angle not
equal to $2\pi /3$. As displayed in Fig~\ref{fig:excitations} (b), the kink
separates the domain wall into two segments. Since the chiralities should be
changed by the permutation of spin orientations within the elementary
triangles when going across the segments, a discrepancy of $2\pi /3$ is
introduced on a string connecting the kinks. The fractional vortices with
the topological charge of $1/3$ thus form at the centers of the hexagons as
the terminals of the string. Hence, the anti-ferromagnetic XY model on the
kagome lattice should go through at least two kinds of phase transitions
associated with the proliferation of the domain walls and the unbinding of
fractional vortex pairs.

\begin{figure}[tbp]
\centering
\includegraphics[width=\linewidth]{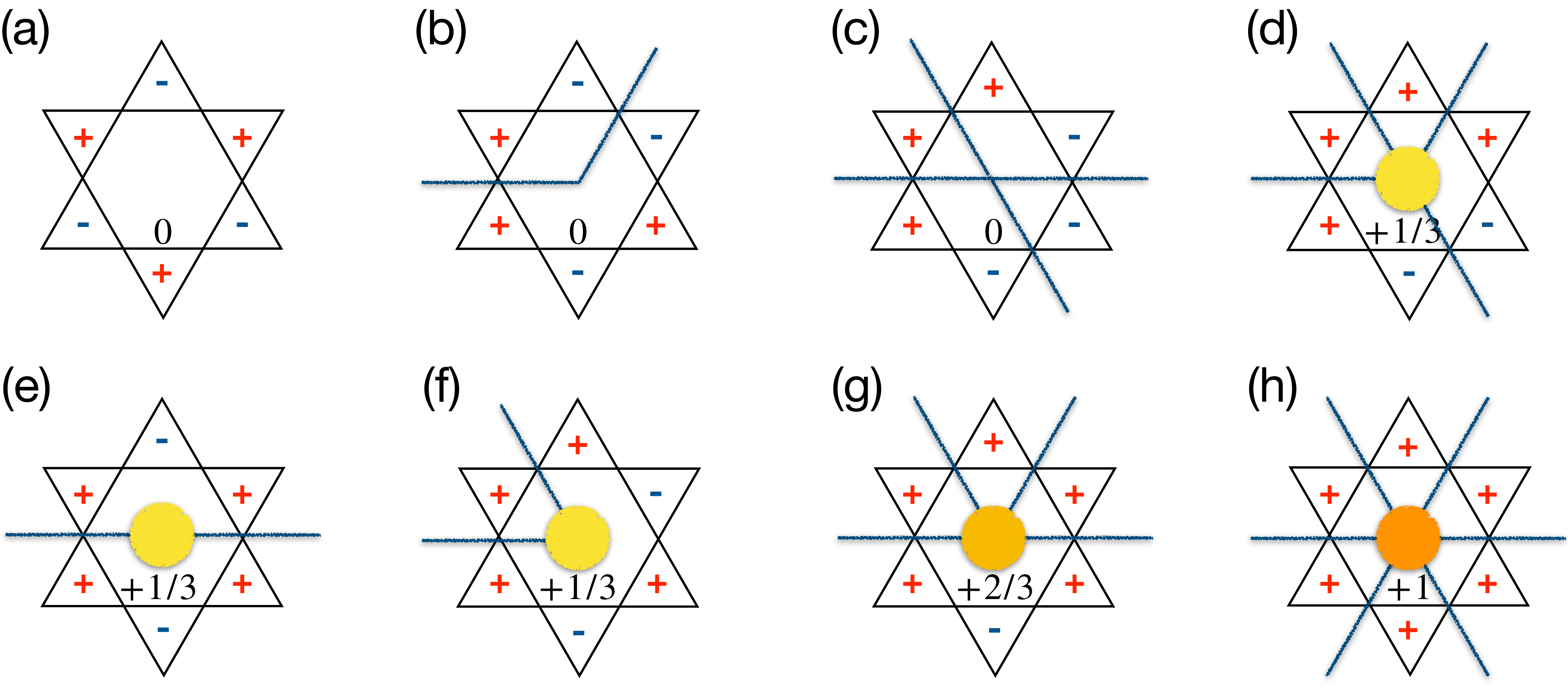}
\caption{ Topological charges of vortices corresponding to different domain
wall structures. Another set of minus vortices can be obtained by inverting
all the signs of chiralities. (a) Regular structure with zero charges. (b)
Zero-energy domain wall with zero charges. (c) Intersection of two straight
walls with zero charges. (d) Intersection of two walls with charge $+\frac{1%
}{3}$. (e) Domain wall of angle $\protect\pi$ with charge $+\frac{1}{3}$.
(f) Domain wall of angle $\frac{\protect\pi}{6}$ with charge $+\frac{1}{3}$.
(g) Intersection of two walls with charge $+\frac{2}{3}$. (h) Intersection
of three straight walls with charge $+1$.}
\label{fig:vortices}
\end{figure}

\begin{figure}[tbp]
\centering
\includegraphics[width=\linewidth]{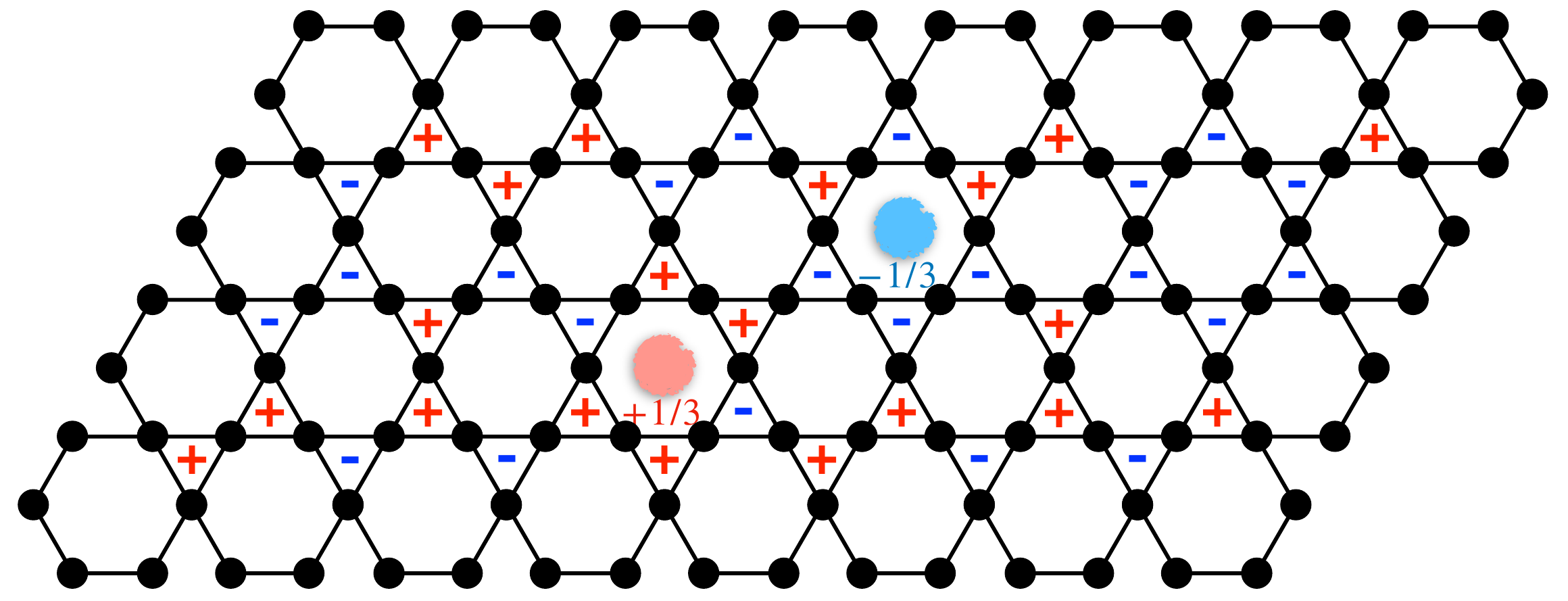}
\caption{ A pair of bound $\pm\frac{1}{3}$ vortices (red and blue circles)
is created by switching the chiralities of two adjacent triangles of one of
the degenerate minimum energy patterns.}
\label{fig:frac13ex}
\end{figure}

On the contrary, some other works\cite%
{Rzchowski_1997,Cherepanov_2001,Park_2001,Andreanov_2020} suggested that the
chirality is always disordered with infinitely many ground states and $\exp
(i3\theta )$ is the lowest-order quantity showing quasi-LRO. Instead of
freezing into one specific pattern, the system would move among all possible
patterns with the equilibrium probability given by the corresponding
Boltzmann weights. Since the domain walls displayed in Fig~\ref%
{fig:excitations} (a) can freely be excited, the physical pictures should be
clearer by introducing the concept of excessive charges\cite{Korshunov_1986}
rather than drawing complicated domain wall configurations. In this way, the
topological charges of vortices on the centers of the hexagons can be
defined by the chiralities of six surrounding triangles
\begin{equation}
q_{h}=\frac{1}{3}\sum_{t=1}^{6}q_{t},
\end{equation}%
where the charges $q_{t}=\pm \frac{1}{2}$ corresponds to chiralities $\sigma
_{t}=\pm 1$, and the factor $\frac{1}{3}$ results from the fact that each
triangle is shared by three neighboring hexagons. The relationships between
the charges of vortices and the local domain wall configurations are
enumerated in Fig~\ref{fig:vortices}. Another set of minus vortices
configurations can be obtained by inverting all the signs of chiralities on
the triangles. It is clear to see that the low temperature phase allows the
free excitations of integer vortices of Fig~\ref{fig:vortices} (a)-(c) and
Fig~\ref{fig:vortices} (h), while the fractional vortices shown in Fig~\ref%
{fig:vortices} (d)-(g) are bound in pairs due to the higher energy cost. As
shown in Fig~\ref{fig:frac13ex}, a pair of $\pm \frac{1}{3}$ vortices is
created by switching the chiralities of two adjacent triangles. As the
temperature increases, the fractional vortices will unbind at the BKT
transition corresponding to the destruction of quasi-LRO in $e^{i3\theta }$.

\section{Tensor network theory}

\subsection{Problems in the standard representation}

The partition function of a classical lattice model with local interactions
can be always represented as a contraction of TN as a product of the
transfer matrices on its original lattice\cite{Zhao_2010} or the dual lattice%
\cite{Levin_2007}. The standard construction of the TN starts by putting a
matrix on each bond accounting for the Boltzmann weight of the neighboring
interactions. Then the local tensors can be obtained from suitable
decompositions of the local bond matrices. Although this paradigm is proven
a success in the studies of the classical ferromagnetic XY model\cite%
{Yu_2014, Vanderstraeten2019_2} and fully frustrated XY model on a square
lattice with careful splits of $U(1)$ spins to encode the ground-state local
rules\cite{Vanderstraeten_2018,Vanhecke_2021,Song_2022,Colbois_2022}, it
cannot be directly applied to the frustrated XY kagome anti-ferromagnets,
where a finite truncation of the interaction matrices fails to include the
massive degeneracy of the low temperature phase.

\begin{figure}[tbp]
\centering
\includegraphics[width=\linewidth]{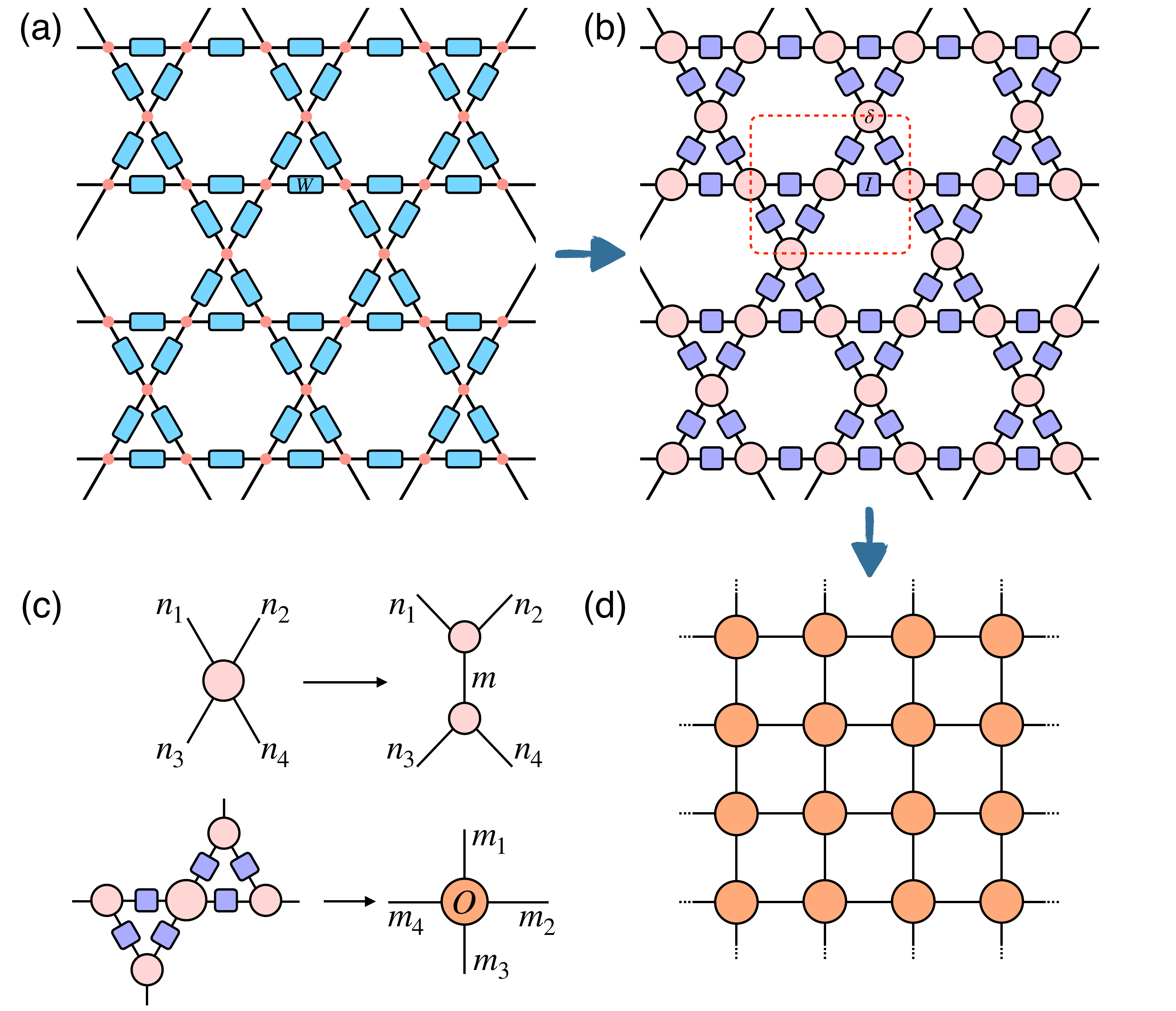}
\caption{ (a) The TN representation of the partition function with
interaction matrices on the links accounting for the Boltzmann weight. (b)
The TN representation of the partition function defined on the original
lattice. The translation invariant cluster is circled by the red dotted
line. (c) The split of the Kronecker delta functions and the construction of
the local $O$ tensor. (d) The TN representation of the partition function
composed of uniform local $O$ tensors.}
\label{fig:kgm_tnerr}
\end{figure}

To illustrate this problem, we first give the TN representation of the
partition function from the generic approach. The partition function on the
original lattice is expressed as
\begin{equation}
Z=\prod_{i}\int \frac{\mathrm{d}\theta _{i}}{2\pi }\prod_{\langle i,j\rangle
}W(\theta _{i},\theta _{j}),
\end{equation}%
where $W(\theta _{i},\theta _{j})=\mathrm{e}^{-\beta J_{1}\cos (\theta
_{i}-\theta _{j})}$ can be viewed as infinite interaction matrices with
continuous $U(1)$ indices, and $\beta =1/k_{B}T$ is the inverse temperature
with $k_{B}$ the Boltzmann constant. The partition function is then cast
into the TN as shown in Fig.~\ref{fig:kgm_tnerr} (a), where the integrations
$\int d\theta _{i}/2\pi $ are denoted as red dots and the matrix indices
take the same values at the joint points.

To transform the local tensors onto a discrete basis, we employ the
character decomposition
\begin{equation}
\mathrm{e}^{x\cos \theta }=\sum_{n=-\infty }^{\infty }I_{n}(x)\mathrm{e}%
^{in\theta }
\end{equation}%
to decompose the interaction matrices as
\begin{equation}
W(\theta _{i},\theta _{j})=\sum_{n_{l}}I_{n_{l}}(-\beta )\mathrm{e}%
^{in_{l}(\theta _{i}-\theta _{j})},
\end{equation}%
where $I_{n}(x)$ are the modified Bessel functions of the first kind. The
index $n_{l}$ lies on the bond $l$ connecting NN sites with a direction from
$i $ to $j$, which means $n_{l}=n_{(i,j)}=-n_{(j,i)}$. After integrating out
the $U(1)$ phase degrees of freedom at each site, we get the partition
function shown in Fig.~\ref{fig:kgm_tnerr} (b) in terms of the TN
representation,
\begin{equation}
Z=\sum_{\{n_{l}\}}\prod_{l}I_{n_{l}}(-\beta )\prod_{i}\delta
_{n_{1}(i)+n_{2}(i)+n_{3}(i)+n_{4}(i),0},  \label{eq:kgm_Z}
\end{equation}%
where the conservation law of $U(1)$ charges has been encoded in the local $%
\delta $ tensors. To obtain a uniform TN and avoid the splits of the
spectrum tensors $I_{n}$ with the negative factor $(-1)^{n}$, we decompose
the $\delta $ tensors, as two neighboring triangles share only one corner.
As shown in Fig.~\ref{fig:kgm_tnerr} (c), we split a bigger 4-leg $\delta $
tensor into two smaller 3-leg $\delta $ tensors
\begin{equation}
n_{1}+n_{2}+n_{3}+n_{4}=0\quad \rightarrow
\begin{cases}
n_{1}+n_{2}+m=0 \\
n_{3}+n_{4}+m=0%
\end{cases}%
.
\end{equation}%
Then we group the inner tensors composing the transitional invariant unit
circled by the red dotted line in Fig.~\ref{fig:kgm_tnerr} (b) into a 4-leg $%
O$ tensor. Finally, the uniform TN displayed in Fig.~\ref{fig:kgm_tnerr} (d)
is obtained as%
\begin{equation}
Z=\mathrm{tTr}\prod_{s}O_{m_{1},m_{2}}^{m_{3},m_{4}}(s),
\end{equation}%
where \textquotedblleft $\mathrm{tTr}$\textquotedblright\ means the tensor
contraction over all auxiliary links and $s$ denotes the sites of the
transitional invariant unit.

Unfortunately, such a construction does not work in the low temperature
regime. Although the standard contraction algorithms converge for the
construction of TN without a hitch, they lead to incorrect contraction
results at low temperatures. The contraction values are found to be highly
dependent on the truncation dimensions of the Boltzmann weight, i.e., the
upper limit $n_{\max }$ of the Bessel function expansions $I_{n}(-\beta )$.
The bond dimension $d$ of the $I$ tensor in Fig.~\ref{fig:kgm_tnerr} (b) is
defined as $d=2n_{\max }+1$, where the index $n$ runs in the range $%
\{-n_{\max },\cdots ,n_{\max }\}$. The free energy density shown in Fig.~\ref%
{fig:fe_err} is calculated directly from the partition function
\begin{equation}
f=-\frac{1}{\beta N}\ln Z,
\end{equation}%
where $N$ is the total number of sites on the original kagome lattice under
contraction. We find that the free energy always displays an inflection
point that shifts left with increasing bond dimensions $d$. The singular
point in the free energy density seems to indicate the occurrence of a
first-order phase transition, which turns out to be misleading seen in the
following sections. The move of the inflection point stops around $T\simeq
0.08J_{1}$ and the data below this temperature displays strong fluctuations
upon increasing $d$ thereafter.
\begin{figure}[tbp]
\centering
\includegraphics[width=\linewidth]{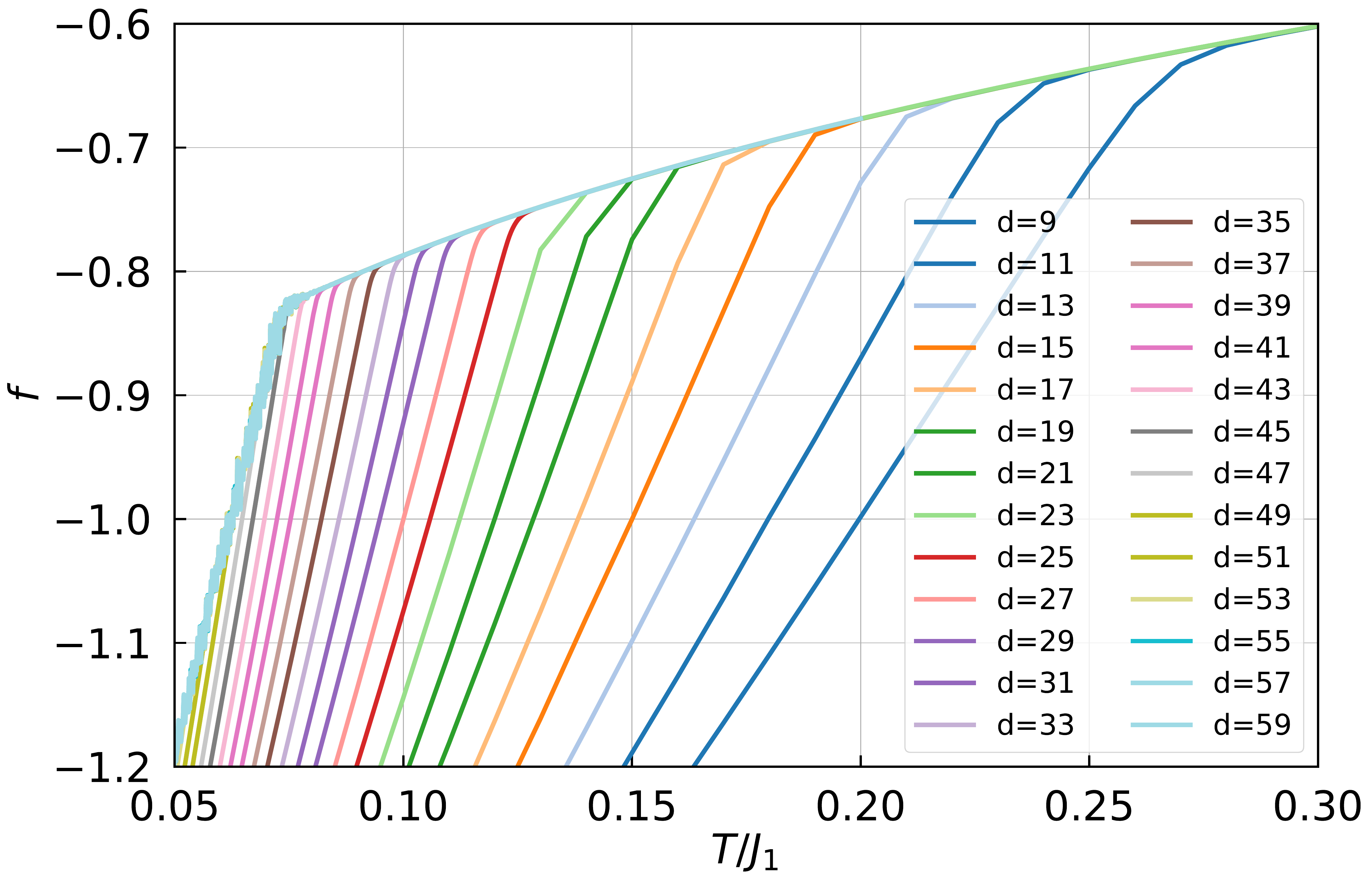}
\caption{ (a) The free energy obtained from the contraction of the TN
constructed from direct truncations on the Boltzmann weight. The bond
dimension $d$ denotes the upper limit of the Bessel function expansions.}
\label{fig:fe_err}
\end{figure}

The low temperature physics should be calculated with infinite bond
dimension $d$ in principle, because the Bessel functions become $I_{n}(\beta
)\gg 1$ with $\beta \gg 1$. However, for numerical calculations, a finite
truncation on $d$ is necessary. The finite cutoff corresponds to the
saturation of a maximal number of NN bonds, which is far away from the true
ground state of infinite degeneracy due to the strong frustrations. That is
why the standard construction is not applicable in the low temperature
regime.

\subsection{Duality transformation}

\begin{figure}[tbp]
\centering
\includegraphics[width=\linewidth]{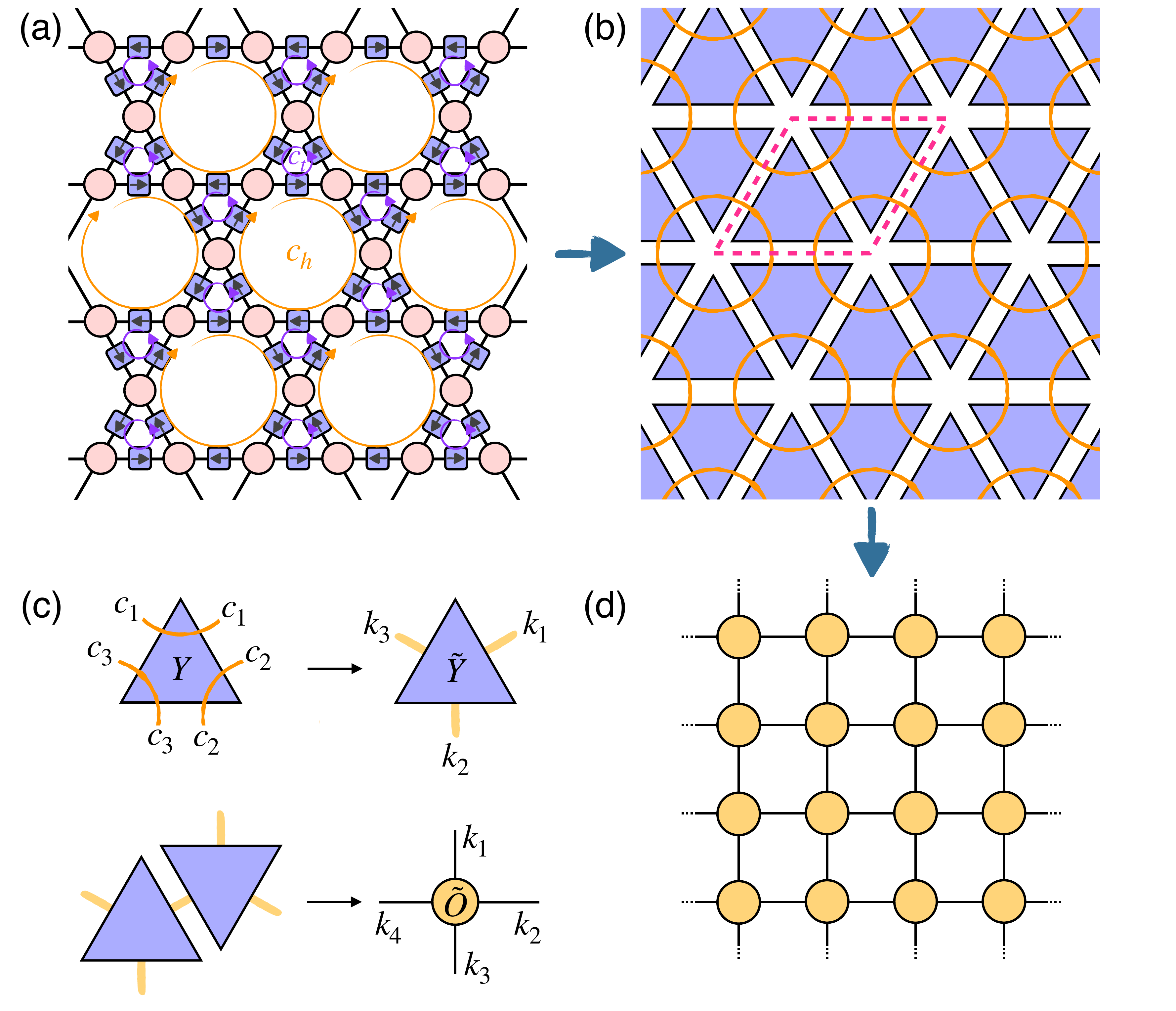}
\caption{ (a) The decompositions of $n_{l}$ at each kagome lattice bond in
terms of integer-valued currents $c_{h}$ and $c_{t}$ circulating on the
centers of elementary hexagons and triangles. (b) The TN consisting of local
$Y$ tensors (purple triangles). The orange circles denote the hexagons
currents $c_{h}$ and the dashed box denotes the transitional invariant unit.
(c) The basis transformation from $Y$ tensor to $\tilde{Y}$ tensor and the
construction of local $\tilde{O}$ tensor by combining two neighboring $%
\tilde{Y}$ tensors. (d) The TN representation of the partition function
composed of uniform local $\tilde{O}$ tensors.}
\label{fig:kgm_tn}
\end{figure}

To avoid the direct truncation on Bessel function expansion $n$, we propose
a new construction approach with the help of the duality transformation\cite%
{Jose_1977,Knops_1977,Cherepanov_2001}. As shown in Fig.~\ref{fig:kgm_tn}
(a), we decompose $n$ in terms of integer-valued currents circulating on the
centers of elementary hexagons and triangles
\begin{equation}
n_{l}=c_{h}+c_{t},
\end{equation}%
where the arrows on the links denote the directions assigned to $n_{l}$. The
negative value of $c_{h}$ or $c_{t}$ means the reverse direction of the
current against $n_l$. In this way, the conservation condition on each site $%
n_{1}+n_{2}+n_{3}+n_{4}=0$ is satisfied automatically. And the partition
function \eqref{eq:kgm_Z} is transformed into
\begin{equation}
Z=\sum_{\{c_{h},c_{t}\}}\prod_{l}I_{c_{h}+c_{t}}(-\beta ).
\end{equation}%
Then the problem with finite truncations of the Bessel functions can be
bypassed by performing an infinite sum over the triangle currents
\begin{equation}
Z=\sum_{\{c_{h}\}}\prod_{t}\sum_{c_{t}=-\infty }^{\infty
}\prod_{h_{t}=1}^{3}I_{c_{h_{t}}+c_{t}}(-\beta ),  \label{eq:Z_th}
\end{equation}%
where $h_{t}=1,2,3$ denote three hexagons surrounding each triangle $t$.
Using the Fourier transformation
\begin{equation}
I_{n}(-\beta )=\frac{1}{2\pi }\int_{-\pi }^{\pi }d\theta \mathrm{e}%
^{-in\theta }\mathrm{e}^{-\beta \cos (\theta )},
\end{equation}%
we sum out the $c_{t}$ on each triangle and obtain

\begin{widetext}
\begin{eqnarray}
Y_{c_{1},c_{2},c_{3}}&&=\sum_{c_{t}=-\infty }^{\infty
}\prod_{h=1}^{3}I_{c_{h}+c_{t}}(-\beta )  \notag \\
&&=\sum_{c_{t}=-\infty }^{\infty } \int\limits_{-\pi}^{\pi}\int\limits_{-\pi}^{\pi}\int\limits_{-\pi}^{\pi} \frac{d\phi _{1}d\phi _{2}d\phi _{3}%
}{(2\pi )^{3}}\mathrm{\exp }\left( -\sum_{h=1}^{3}[i(c_{h}+c_{t})\phi
_{h}+\beta \cos \phi _{h}]\right)  \notag \\
&&=\int\limits_{-\pi}^{\pi}\int\limits_{-\pi}^{\pi} \frac{d\phi _{1}d\phi _{2}}{(2\pi )^{2}}\exp \left( -i[\phi
_{1}(c_{1}-c_{3})+\phi _{2}(c_{2}-c_{3})]-\beta \lbrack \cos \phi _{1}+\cos \phi
_{2}+\cos (\phi _{1}+\phi _{2})]\right) .
\label{eq:YJ}
\end{eqnarray}%
Here $Y_{c_{1},c_{2},c_{3}}$ is referred to a 6-rank tensor with indices $%
c_{1}$, $c_{2}$ and $c_{3}$ displayed in Fig.~\ref{fig:kgm_tn} (b), where
the neighboring indices at each corner of the triangle share the same $c_{h}$.

Since the transition temperature is quite low\cite{Rzchowski_1997}, the
integrand in \eqref{eq:YJ} is saturated by the vicinity of the saddle points
$\phi _{h}\sim \frac{2\pi \sigma _{t}}{3}$ in the low temperature limit ($%
\beta \gg 1$), where $\sigma _{t}=\pm 1$ is the chirality defined on the
triangles related to infinite ground states. Substituting the asymptotic formula
\begin{eqnarray}
Y_{c_{1},c_{2},c_{3}} &\propto &\sum_{\sigma _{t}=\pm 1} \int\limits_{-\pi}^{\pi}\int\limits_{-\pi}^{\pi} \frac{d\phi
_{1}d\phi _{2}}{(2\pi )^{2}}\exp \left\{ -i[\phi _{1}(c_{1}-c_{3})+\phi
_{2}(c_{2}-c_{3})]-\frac{\beta }{2}[(\phi _{1}-\frac{2\pi \sigma
_{t}}{3})^{2}+(\phi _{2}-\frac{2\pi \sigma _{t}}{3})^{2}]\right\}   \notag \\
&=&\sum_{\sigma _{t}=\pm 1}\exp \left\{ -\frac{i2\pi \sigma _{t}}{3}%
(c_{1}+c_{2}+c_{3}) - \frac{1}{3\beta }
[(c_{1}-c_{3})^{2}+(c_{2}-c_{3})^{2}+(c_{3}-c_{1})^{2}]\right\},
\label{eq:Y_dc}
\end{eqnarray}%
back into the expression \eqref{eq:Z_th} and using the Poisson summation formula
\begin{equation}
\sum_{c=-\infty }^{\infty }\delta (\tilde{c}-c)=\sum_{m=-\infty }^{\infty }%
\mathrm{e}^{i2\pi m\tilde{c}},
\end{equation}%
we can write the partition function as
\begin{equation}
Z=\sum_{\{\sigma _{t}\},\{m_{h}\}}\left( \prod_{h}\int\limits_{-\infty}^{\infty} d\tilde{c}_{h}\right)
\prod_{h}\exp\left[-i2\pi\sum_{t_{h}=1}^{6} (\frac{\sigma _{t_{h}}}{3}%
+m_{h})\tilde{c}_{h}-\frac{1}{3\beta }\sum_{\langle h,h^{\prime }\rangle }%
(\tilde{c}_{h}-\tilde{c}_{h^{\prime }})^{2}\right],
\end{equation}%
\end{widetext}
where $\tilde{c}$ are continuous currents and $m$ are integers defined on
the centers of the hexagons. In the low temperature limit, the integration
over $\tilde{c}$ gives the conservation conditions of chiralities
surrounding each hexagon
\begin{equation}
\sum_{t_h=1}^{6}\sigma _{t_h}\equiv 0\pmod{6},
\end{equation}%
corresponding to the free excitations of integer vortices as shown in Fig~%
\ref{fig:vortices} (a)-(c) and Fig~\ref{fig:vortices} (h).

Since the values of $Y_{c_{1},c_{2},c_{3}}$ in \eqref{eq:YJ} are just
dependent on the difference of $c_{h}$ on the edges of the triangle, we can
transform the $Y$ tensors onto a new basis
\begin{equation}
k_{1}=c_{1}-c_{2},\quad k_{2}=c_{2}-c_{3},\quad k_{3}=c_{3}-c_{1},
\end{equation}%
and obtain the 3-rank local tensors
\begin{equation}
\tilde{Y}_{k_{1},k_{2},k_{3}}=Y_{c_1,c_2,c_3}
\end{equation}
as shown in Fig.~\ref{fig:kgm_tn} (c). Finally, the transitional invariant
local tensor $\tilde{O}$ is achieved by combining a pair of up and down
triangles. And the uniform TN representation of the partition function
displayed in Fig.~\ref{fig:kgm_tn} (d) is given by
\begin{equation}
Z=\mathrm{tTr}\prod_{s}\tilde{O}_{k_{1},k_{2}}^{k_{3},k_{4}}(s).
\end{equation}

From the asymptotic form \eqref{eq:Y_dc}, it is clear that the tensor
components of $\tilde{Y}$ decrease exponentially with increasing $|k_{i}|$ ($%
i=1,2,3$). Hence we can truncate the series safely and approximate $\tilde{O}
$ with a finite bond dimension $\tilde{d}$ in high precision, where $\tilde{d%
}=2k_{\max}+1$ with the index $k$ ranging from $-k_{\max}$ to $k_{\max}$.

\subsection{Evaluations of the physical quantities}

\begin{figure}[tbp]
\centering
\includegraphics[width=\linewidth]{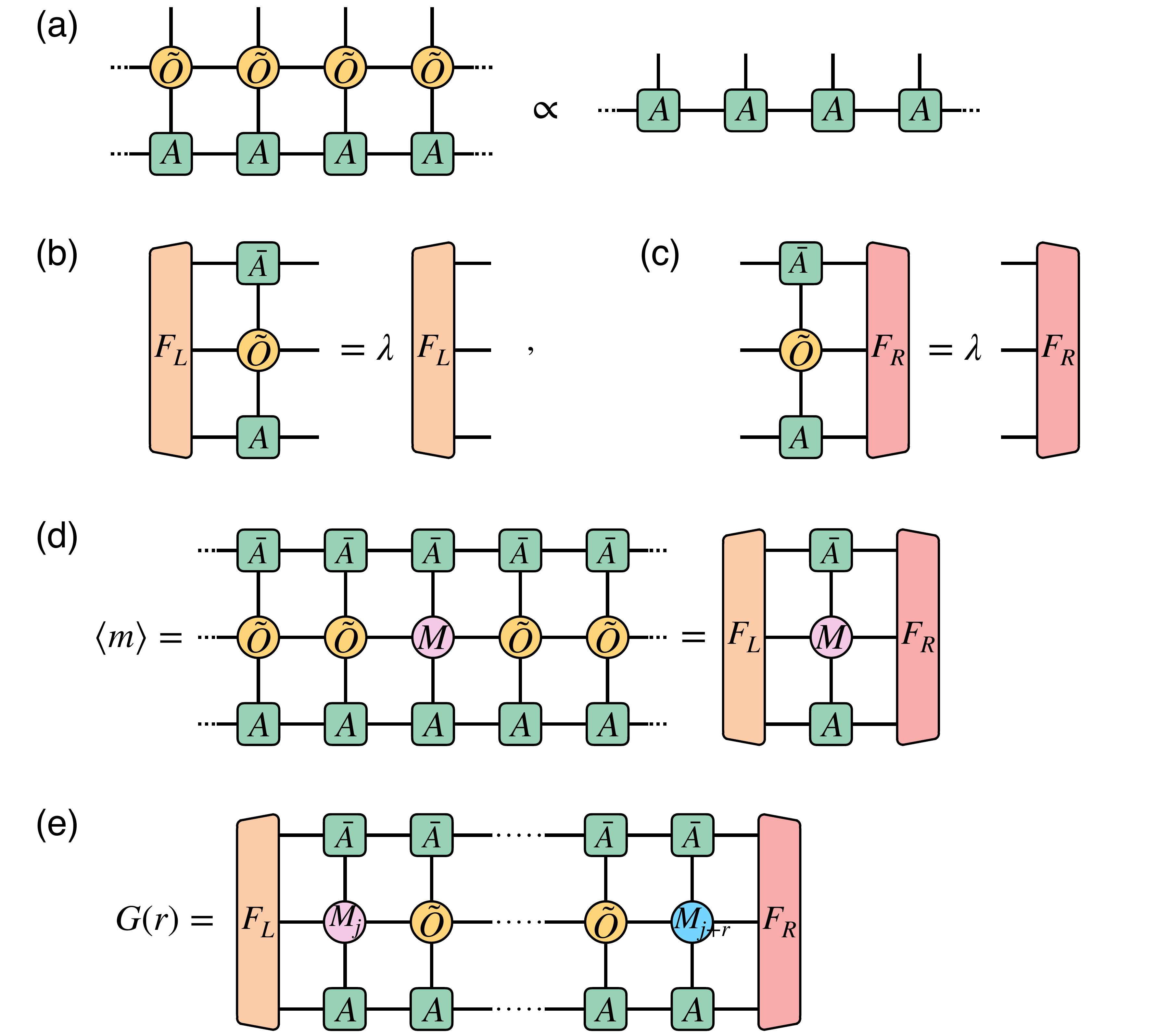}
\caption{ (a) Eigenequation for the fixed-point uniform MPS $|\Psi
(A)\rangle $ of the 1D transfer operator $\hat{T}$. (b) and (c)
Eigenequations for the left and right fixed-point eigenvectors of the
channel operator $\mathbb{T}_{\tilde{O}}$. (d) Expectation of a local
operator calculated by contracting the leading eigenvectors of the channel
operators. (e) Two-point correlation function represented by contracting a
train of channel operators.}
\label{fig:vumps}
\end{figure}

The fundamental object for the calculation of the partition function is the
row-to-row transfer operator consisting of an infinite train of $\tilde{O}$
tensors
\begin{equation}
\hat{T}(\beta )=\sum_{\cdots ,s,p,q,\cdots }\mathrm{Tr}\,\left[ {\cdots
\tilde{O}_{k_{1}}^{k_{3}}(s)\tilde{O}_{k_{1}}^{k_{3}}(p)\tilde{O}%
_{k_{1}}^{k_{3}}(q)\cdots }\right] ,
\end{equation}%
where $s,\ p,\ q,\ \cdots $ label different sites within a single row. This
operator can be regarded as the matrix product operator (MPO) for 1D quantum
spin chains of complicated interactions with the correspondence
\begin{equation}
\hat{H}_{\text{1D}}=-\frac{1}{\beta }\log \hat{T}(\beta ).
\end{equation}%
In the thermodynamic limit, the partition function is determined by the
dominant eigenvalue of the transfer operator as shown in Fig.~\ref{fig:vumps}
(a)
\begin{equation}
\hat{T}(\beta )|\Psi (A)\rangle =\Lambda _{\max }|\Psi (A)\rangle ,
\end{equation}%
where $|\Psi (A)\rangle $ is the leading eigenvector represented by matrix
product states (MPS) made up of uniform local $A$ tensors\cite%
{Zauner_Stauber_2018}. The local tensor $A_{\alpha ,\beta }^{k}$ is a 3-leg
tensor with physical bond dimension $\tilde{d}$ and auxiliary bond dimension
$D$ controlling the accuracy of the approximation.

The fixed-point eigen-equation can be accurately solved by the variational
uniform matrix product state (VUMPS) algorithm\cite%
{Zauner_Stauber_2018,Fishman_2018,Vanderstraeten_2019}, which provides an
efficient variational scheme to approximate the largest eigenvector $|\Psi
(A)\rangle $. During the iteration of a set of optimized eigen-solvers, we
also obtain the leading left and right eigenvectors of the channel
operators. The channel operators have a sandwich structure composed of two
local tensors of fixed point MPS and the middle four-leg local tensor. The
channel operator related to $A$ tensor is defined by
\begin{equation}
\mathbb{T}_{X}=\sum_{i,j}\bar{A}^{i}\otimes X^{i,j}\otimes A^{j},
\end{equation}%
and other channel operators are defined in a similar way. The left and right
fixed points of the channel operator $\mathbb{T}_{\tilde{O}}$ is obtained
from the eigen-equations
\begin{equation}
\left\langle F_{L}\right\vert \mathbb{T}_{\tilde{O}}=\lambda \left\langle
F_{L}\right\vert ,\qquad \mathbb{T}_{\tilde{O}}\left\vert F_{R}\right\rangle
=\lambda \left\vert F_{R}\right\rangle ,  \label{eq:channel_eigen}
\end{equation}%
as displayed in Fig.~\ref{fig:vumps} (b) and (c).

Once the fixed points are achieved, various physical quantities can be
accurately calculated in the TN language. The entanglement properties can be
detected via the Schmidt decomposition of $|\Psi(A)\rangle$ which bipartites
the relevant 1D quantum state of the MPO,
\begin{equation}
|\Psi(A)\rangle=\sum_{\alpha,\beta=1}^{D}s_{\alpha}\delta_{\alpha,\beta}|%
\Psi_\alpha^{-\infty,n}\rangle|\Psi_\beta^{n+1,\infty}\rangle.
\end{equation}
And the entanglement entropy\cite{Vidal_2003} is determined directly from
the singular values $s_\alpha$ as
\begin{equation}
S_E = -\sum_{\alpha=1}^{D}s_\alpha^2\ln s_{\alpha}^2,  \label{eq:ee}
\end{equation}
in correspondence to the quantum entanglement measure.

Moreover, the expectation value of a local observable $m(\theta _{i})$ can
be expressed as
\begin{equation}
m(\theta _{i})=\frac{1}{Z}\prod_{s}\int \frac{d\theta _{s}}{2\pi }\mathrm{e}%
^{-\beta E(\{\theta _{s}\})}m(\theta _{i}),
\end{equation}%
where $E(\{\theta _{s}\})$ is the energy of the state under a given spin
configuration $\{\theta _{s}\}$. For the observables in the form of $%
m(\theta _{i})=\mathrm{e}^{iq\theta _{i}}$, it can be calculated by
inserting the corresponding impurity tensors into the original tensor
network for the partition function. The impurity tensor $%
M_{k_{1},k_{2}}^{k_{3},k_{4}}$ for the observable can be simply constructed
by changing the corresponding charge conservation condition at site $i$ into
\begin{equation}
\sum_{l=1}^{4}n_{l}=q,
\end{equation}%
which introduces imbalanced currents in the local tensor $\tilde{Y}$ with
the mapping $\tilde{Y}_{k_{1},k_{2},k_{3}}\rightarrow \tilde{Y}%
_{k_{1}+q,k_{2},k_{3}}$. Using the MPS fixed point, the contraction of the
TN containing the impurity tensor is reduced to a trace of an infinite
sequence of channel operators,
\begin{equation}
\langle m(\theta )\rangle =\mathrm{Tr}\,\left( \cdots \mathbb{T}_{\tilde{O}}%
\mathbb{T}_{\tilde{O}}\mathbb{T}_{M}\mathbb{T}_{\tilde{O}}\mathbb{T}_{\tilde{%
O}}\cdots \right) ,
\end{equation}%
which can be further squeezed into a contraction of a small network as shown
in Fig.~\ref{fig:vumps}(d),
\begin{equation}
\langle m(\theta )\rangle =\langle F_{L}|\mathbb{T}_{M}|F_{R}\rangle
\end{equation}%
with the help of leading eigenvectors $\langle F_{L}|$ and $|F_{R}\rangle $.

In the same way, the two-point correlation function between local
observables defined by $G(r)=\langle m(\theta_j) m(\theta_{j+r})\rangle$ can
be evaluated by inserting two local impurity tensors into the original TN.
For the case of the spin-spin correlation function, the evaluation of $%
G(r)=\langle \mathrm{e}^{iq(\theta_{j}-\theta_{j+r})}\rangle$ is reduced to
a trace of a row of channel operators containing two impurity tensors as
shown in Fig.~\ref{fig:vumps} (e)
\begin{equation}
G(r)=\langle F_L|\mathbb{T}_{M_j}\underbrace{\mathbb{T}_{O}\cdots \mathbb{T}%
_{O}}_{r-1}\mathbb{T}_{M_{j+r}}|F_R\rangle.  \label{eq:G}
\end{equation}

\section{Numerical Results}

In the previous sampling approaches the phase transitions were determined by
some kinds of order parameter like the magnetization or Binder cumulent,
where the corresponding numerical results of the simulations displayed a
strong dependence on the system size\cite{Rzchowski_1997,Andreanov_2020} and
a slow convergence under large frustrations\cite%
{Swendsen_1987,Wolff_1989,Park_2001}. Within the TN framework, we bring
modern concepts of quantum entanglement to the frustrated classical XY model
on the kagome lattices by mapping the transfer matrix to a 1D quantum
transfer operator. The entanglement entropy of the fixed-point MPS for the
1D quantum correspondence exhibits singularity at the critical temperatures,
which offers a sharp criterion to determine all possible phase transitions
in the thermodynamic limit, especially for the systems possessing $U(1)$ and
$Z_{2}$ degrees of freedom\cite{Song_2021, Song_2022, Song_2022_2}.

\begin{figure}[tbp]
\centering
\includegraphics[width=\linewidth]{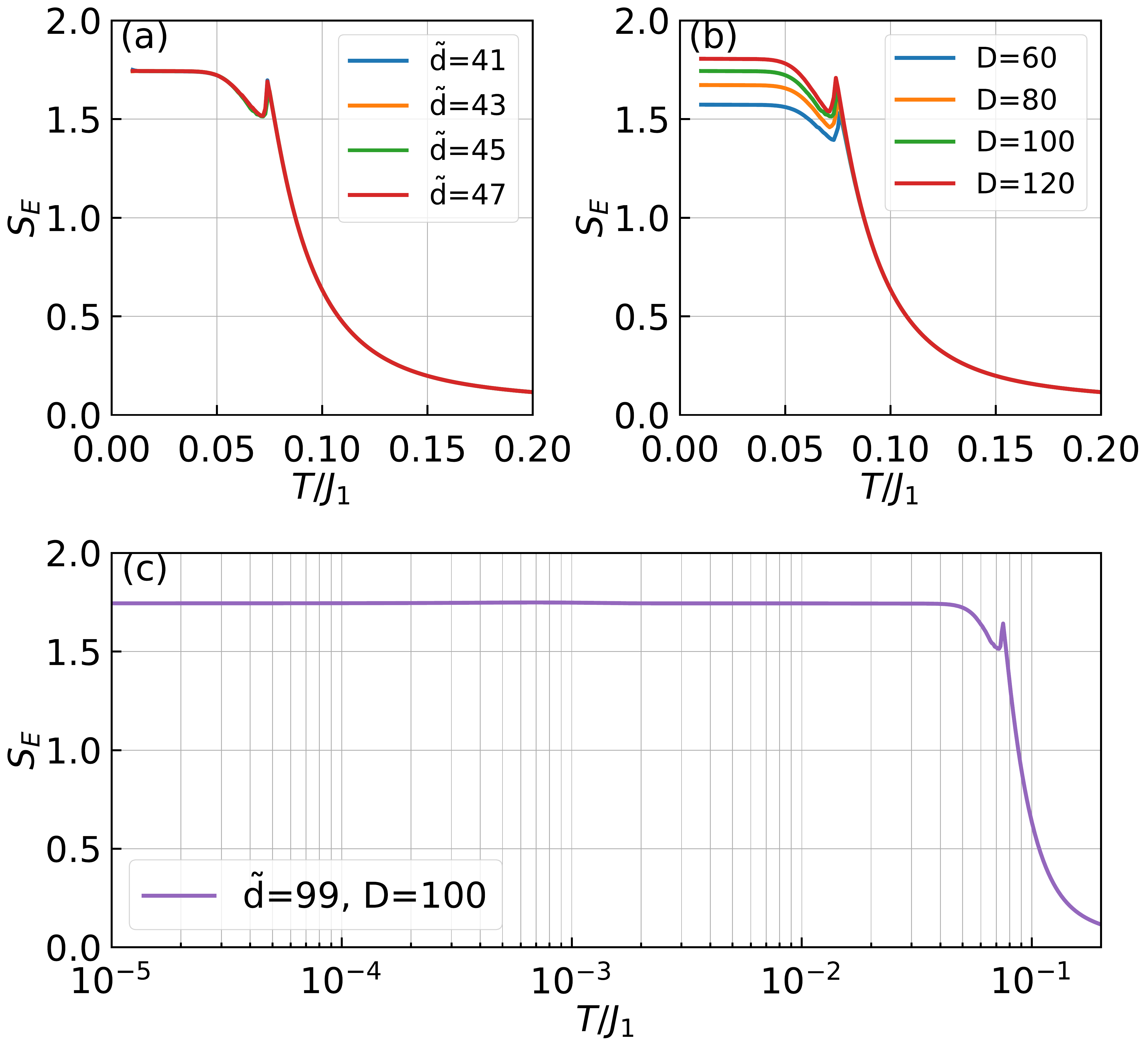}
\caption{ (a) The entanglement entropy as a function of temperature under
MPS bond dimension $D=100$ and different MPO bond dimensions. (b) The
entanglement entropy as a function of temperature under MPO bond dimension $%
\tilde{d}=45$ and different MPS bond dimensions. (c) The entanglement
entropy at low temperatures under MPO bond dimension $\tilde{d}=99$ and MPS
bond dimensions $D=100$.}
\label{fig:kgm_ee}
\end{figure}

As shown in Fig.~\ref{fig:kgm_ee}, the entanglement entropy $S_{E}$ develops
only one sharp singularity at the critical temperature $T_{c}\simeq
0.075J_{1}$, which strongly indicates that a single phase transition takes
place at a rather low temperature. As displayed in Fig.~\ref{fig:kgm_ee}
(a), the entanglement entropies remain unchanged under different bond
dimensions of the local $\tilde{O}$ tensor from $\tilde{d}=41$ to $47$. The
convergence of the entanglement entropy demonstrates the success of the new
construction approach. Moreover, the peak positions are almost unchanged
with different MPS bond dimensions from $D=60$ to $120$ as shown in Fig.~\ref%
{fig:kgm_ee} (b). So we can accurately locate the transition temperature,
which is in good agreement with theoretical expectations\cite%
{Cherepanov_2001,Korshunov_2002} for the unbinding temperature of $1/3$
vortex pairs
\begin{equation}
T_{c}\approx \frac{\pi \sqrt{3}}{72}J_{1}\approx 0.075J_{1}.
\end{equation}%
Since the calculations are performed in the thermodynamic limit, we directly
determine the transition temperature of the finite-size interpolation in the
previous studies using sampling methods\cite{Rzchowski_1997}. The specific
advantage of the TN approach enables us to efficiently investigate the
extremely low temperature regime that has never been reached before at a
small cost of increasing MPO bond dimensions. As shown in Fig.~\ref%
{fig:kgm_ee} (c), the entanglement entropy is smooth everywhere at low
temperatures, which demonstrates that there is no evidence for the selection
of a single ground state of chirality down to $T\sim 10^{-5}J_{1}$. The
temperature has reached low enough to rule out the possibility of
ordering in staggered chirality, where the lower boundary of the transition
temperature was just estimated to $T\sim 10^{-4}J_{1}$\cite{Korshunov_2002}.

For the sake of the possible periodic chirality pattern of $\sqrt{3}\times%
\sqrt{3}$ states, we also build the transfer operator with a larger unit
cell consisting of $3\times3$ clusters of $\tilde{O}$ tensors. The
eigen-equation for the enlarged translational unit can be solved efficiently
by the multiple lattice-site VUMPS algorithm\cite{Nietner_2020}. We find
that the lattice symmetry is not spontaneously broken and all the results
are the same as the case of the simplest unit cell, which means the absence
of the LRO of chirality.

\begin{figure}[tbp]
\centering
\includegraphics[width=\linewidth]{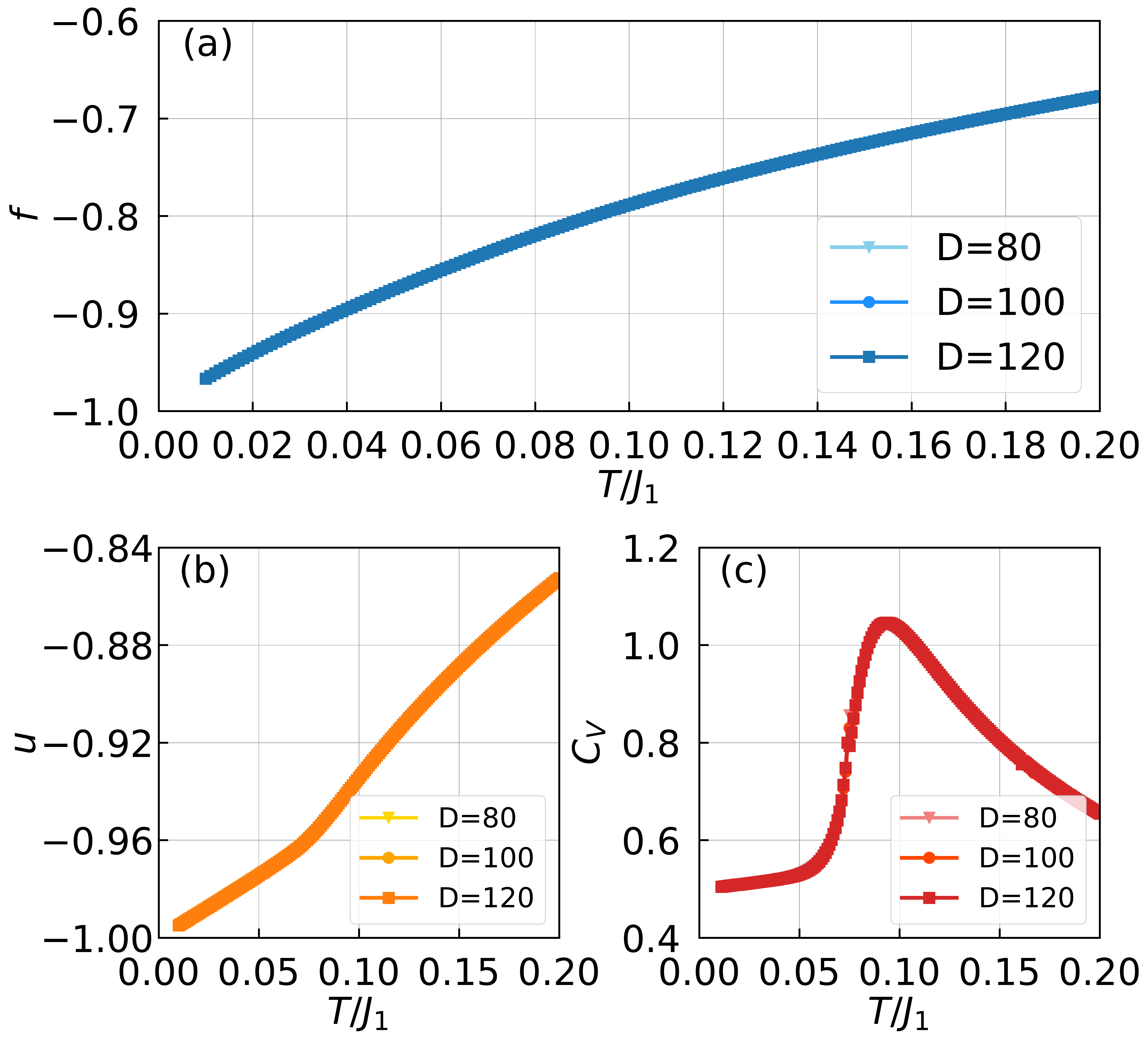}
\caption{ The thermodynamic properties under different MPS bond dimensions.
(a) The free energy as a function of temperature. (b) The internal energy as
a function of temperature. (c) The specific heat as a function of
temperature.}
\label{fig:fucv}
\end{figure}

Furthermore, we study the thermodynamic properties to understand the
essential physics of the phase transition. The free energy per site can be
calculated straightforwardly from the variational MPS setup
\begin{equation}
f=-\frac{1}{3\beta }\ln \lambda ,
\end{equation}%
where $\lambda $ is the eigenvalue of the channel operators in %
\eqref{eq:channel_eigen} and the coefficient $\frac{1}{3}$ is due to the
fact that each $\tilde{O}$ tensor contains three original kagome lattice
sites. As displayed in Fig.~\ref{fig:fucv} (a), it is clear that the free
energy shows no signs of a first-order phase transition as it is perfectly
smooth everywhere. The sharp contrast to the free energy in Fig.~\ref%
{fig:fe_err} obtained from the standard construction of TN demonstrates the
success of our new construction approach.

The internal energy density can be obtained from the real part of the
expectation value of the nearest-neighbor two-angle observable
\begin{equation}
u=2\langle \cos (\theta _{i}-\theta _{i+1})\rangle =2\mathrm{Re}\langle
e^{i(\theta _{i}-\theta _{i+1})}\rangle \mathrm{.}
\end{equation}%
As shown in Fig.~\ref{fig:fucv} (b), we find that there is no singularity in
the temperature dependence of the internal energy. As the temperature
approaches $T\rightarrow 0$, the internal energy converges to $%
u(T)\rightarrow -1$, which implies that the angle between the NN spins
becomes $2\pi /3$ at the ground state. Also, the expectation value of
chirality can be calculated as
\begin{equation}
\langle \sigma \rangle =\frac{2}{3\sqrt{3}}\sum_{\langle i,j\rangle \in
\triangle }\sin (\theta _{i}-\theta _{j})
\end{equation}%
in the same way as the frustrated triangular lattices\cite{Miyashita_1984}.
We find the expectation value of chirality equals zero with $\mathrm{Im}%
\langle e^{i(\theta _{i}-\theta _{i+1})}\rangle =0$ at all temperatures,
which means the presence of strong fluctuations of chirality. Furthermore,
the specific heat can be derived directly from
\begin{equation}
C_{V}=\frac{du}{dT},
\end{equation}%
which develops a round bump around $T=0.095J_1$ higher than the transition
temperature determined from the entanglement entropy, which is a
characteristic feature for the BKT transition. The smooth behavior of
thermodynamic properties rules out the possibility of the first-order
transition from the lifting of chirality degeneracy\cite{Korshunov_2002}.

\begin{figure}[tbp]
\centering
\includegraphics[width=\linewidth]{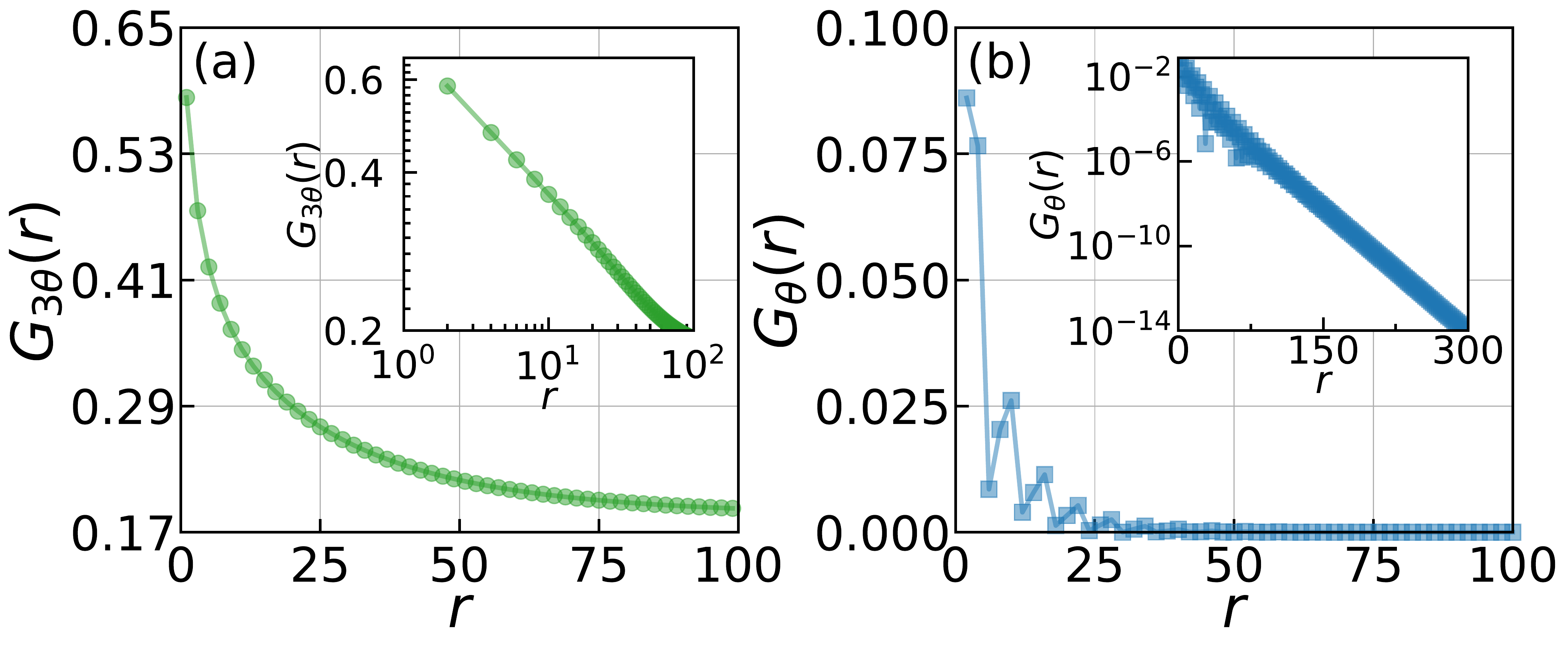}
\caption{ A comparison between different correlation functions at $T=0.07J_1$
below the BKT transition. (a) The correlation function of the $1/3 $
fractional vortices exhibits a power law decay. (b) The correlation function
of the integer vortices shows an exponential decay.}
\label{fig:clfn}
\end{figure}

To further explore the nature of the phase transition, we calculate the
following two different correlation functions among integer vortices and
fractional vortices
\begin{align}
G_{\theta }(r)& =\langle \cos (\theta _{j}-\theta _{j+r})\rangle,  \notag \\
G_{3\theta }(r)& =\langle \cos (3\theta _{j}-3\theta _{j+r})\rangle .
\end{align}
A comparison between these two correlation functions in the low temperature
phase is shown in Fig.~\ref{fig:clfn}. For a given temperature of $%
T=0.07J_1<T_{c}$, the correlation function $G_{3\theta }(r)$ displays a
power law behavior with distance but the correlation function $G_{\theta
}(r) $ decays exponentially, corresponding to the presence of quasi-LRO of
the fractional vortices and anti-vortices. As the temperature increases
above $T_{c}$, both correlation functions behave in an exponential way,
indicating that the system goes into the disordered phase. It is interesting
to see that the correlation function $G_{\theta }(r)$ displays a dumped
oscillation in the short range in accordance with the $\sqrt{3}\times \sqrt{3%
}$ pattern, but the fluctuation amplitude decays rapidly as shown in Fig.~%
\ref{fig:clfn} (b). The exponential behavior in $\theta $ fields can be
explained by the free fluctuations of the system among different
ground-state chirality patterns. Although the spin wave fluctuations favor
the short-range anti-ferromagnetic arrangement of chiralities, they are too
weak to lift the ground-state degeneracy. The uncertainty of the phase
difference of $\pm 2\pi /3$ increases with distance, which destroys the
correlation in $\theta $ fields in the long range.

Such a behavior clearly indicates that the integer vortices can always
excite freely, implying the absence of phase coherence between Cooper pairs
at large distances. However, the $1/3$ fractional vortex pairs with
quasi-LRO survive at low temperatures, which can be regarded as the
condensation of Cooper sextuples. This phenomenon is just the characteristic
of the charge-6e superconductivity\cite{Ge_2022}. As the temperature
increases, the system goes into the normal phase through a BKT transition
driven by the dissociations of $1/3$ vortices.

\section{Conclusion and Outlook}

\begin{figure}[tbp]
\centering
\includegraphics[width=\linewidth]{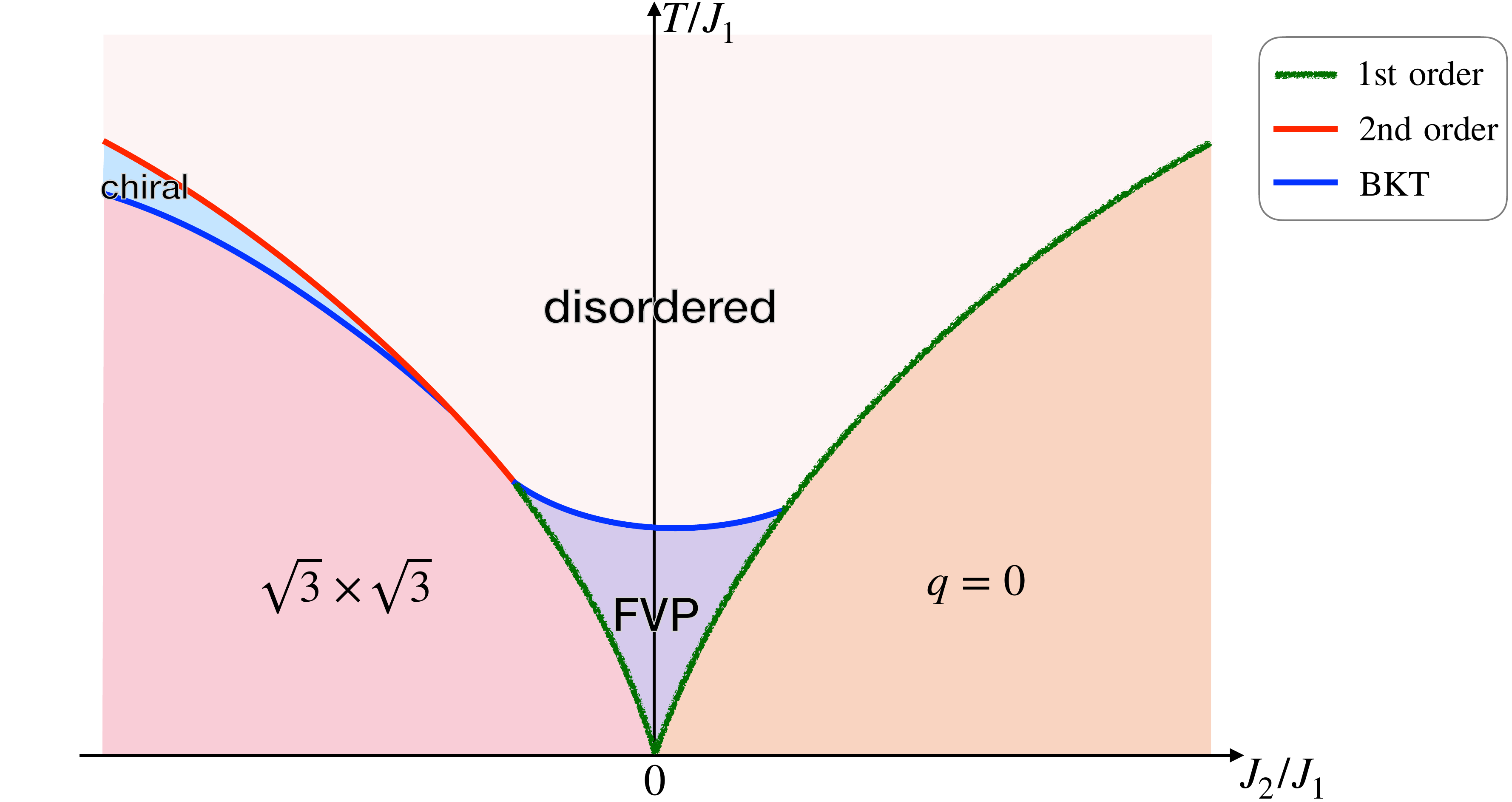}
\caption{ The schematic global phase diagram of the antiferromagnetic XY
model with both NN and NNN interactions on kagome lattices defined by \eqref{eq:NNN_H}. The $q=0
$ and $\protect\sqrt{3}\times \protect\sqrt{3}$ ordered states denote the
LRO of the ferromagnetic and anti-ferromagnetic chirality pattern
respectively, accompanying the quasi-LRO in the integer vortices. The FVP is
the abbreviation for the fractional vortex-antivortex paired phase, where
the $1/3$ fractional vortices are bounded in pairs in the absence of integer
vortex-antivortex pairing. In the chiral phase, the phase coherence of
either integer vortices or fractional vortices are destroyed while the
anti-ferromagnetic chiral LRO survives.}
\label{fig:phases}
\end{figure}

Using the state-of-the-art TN methods, we have clarified the nature of the
critical phase and phase transition of the fully frustrated classical XY
model on 2D kagome lattice, which provides a plausible origin to the
charge-6e SC discovered recently\cite{Ge_2022}. We find standard
construction approach of the TN for the partition function does not work due
to the strong frustrations of the kagome lattice. To solve the issue of
direct truncation of the Boltzmann weight, we have introduced a new method
to build the TN based on duality transformation. Then the partition function
is expressed as a product of 1D transfer matrix operators, whose
eigen-equation is solved by the VUMPS algorithm accurately. The singularity
of the entanglement entropy for this 1D quantum correspondence provides a
stringent criterion for the temperature of possible phase transitions.
Through a systematic analysis of thermodynamic properties and correlation
functions in the thermodynamic limit, a single BKT transition is confirmed, 
which is driven by the unbinding of $1/3$ fractional vortex-antivortex pairs 
at $T_{c}\simeq 0.075J_{1}$. The absence of LRO of chirality or phase coherence 
between integer vortices is verified at all temperatures. Thus the long 
standing controversy about the phase transitions in the fully frustrated 
XY spin model on a kagome lattice is solved. The low temperature phase of the
model can be interpreted as the presence of charge-6e SC but in the absence
of the charge-2e SC.

An interesting open problem is to determine the full phase diagram with both 
the NN and NNN interactions described by \eqref{eq:NNN_H}. For both signs of 
the NNN interaction, the degeneracy of the ground state is lifted to 
$U(1)\times Z_{2}$. The possible global phase diagram is schematically described 
in Fig.~\ref{fig:phases}. At low temperatures, the anti-ferromagnetic NNN
interaction ($J_{2}>0$) will drive the system into the $q=0$ states with
uniform chirality, while the ferromagnetic NNN interaction ($J_{2}<0$)
drives the system into the $\sqrt{3}\times \sqrt{3}$ ordered state with a
finite staggered chirality. The LRO of chirality allows the quasi-LRO of the
integer vortices corresponding to the phase coherence of charge-2e SC. The 
phase transitions between the charge-2e SC and charge-6e SC states could be 
driven by proliferations of the low energy domain walls, which was supposed to 
be a first-order phase transition (the green lines) described by a six-state 
model\cite{Korshunov_2002}. And the phase boundary between the charge-6e SC 
state and the normal state still belongs to BKT transition (the blue line) 
driven by unbinding of $1/3$ fractional vortex-antivortex pairs. From the 
studies on the frustrated XY model on a triangular or square 
lattice\cite{Lee_1998,Capriotti_1998,Song_2022}, we can expect that
the staggered chiral LRO may survive above the BKT transition (the blue
line) with an intermediate a chiral ordered phase. In this way, the boundary
between the chiral ordered phase and disordered phase should be a second-order 
phase transition (the red line). We believe that our TN approach should provide 
a promising way to further investigate the fully frustrated XY on the kagome 
lattice with both NN and NNN interactions, giving rise to more interesting 
and fruitful insights into the exotic phenomena in the kagome superconducting 
materials.

\begin{acknowledgments}
The research is supported by the National Key Research and Development
Program of MOST of China (2016YFYA0300300 and 2017YFA0302902).
\end{acknowledgments}

\bibliography{reference}

\end{document}